\documentclass{aastex631}

\def\mnras{Mon. Not. Roy. Astr. Soc.}
\def\prl{Phys. Rev. Lett.}
\def\prc{Phys. Rev. C}

\def\apjl{Astrophys. J. Lett.}

\def\eg{{\it e.\thinspace g.}}

\def\ie{{\it i.\thinspace e.}}
\def\gcc{\hbox{\rm\hskip.35em  g cm}$^{-3}$}
\def\cms{\hbox{\rm\hskip.35em  cm s}$^{-1}$}

\DeclareMathAlphabet\mathbfcal{OMS}{cmsy}{b}{n}

\newcommand{\half}{\frac{1}{2}}

\newcommand{\fbarbf}         {\mbox{\boldmath$\bar{f}$}}
\newcommand{\ubf}         {\mbox{\boldmath$u$}}

\newcommand{\kappabf}         {\mbox{\boldmath$\kappa$}}

\newcommand{\fbf}         {\mbox{\boldmath$f$}}

\newcommand{\vbf}         {\mbox{\boldmath$v$}}

\newcommand{\rbf}         {\mbox{\boldmath$r$}}

\def\half{\frac{1}{2}}

\newcommand{\be}{\begin{equation}}
\newcommand{\ee}{\end{equation}}

\newcommand{\stack}[2]{\ensuremath{
\begin{array}{c}
#1 \\
#2 
\end{array}
}}

\usepackage{graphicx}
\usepackage{float}
\usepackage{hyperref}

\begin{document}

\title{Vortex Pinning in Neutron Stars, Slip-stick Dynamics, and the
 Origin of Spin Glitches}

\author{Bennett Link}
\affiliation{Department of Physics, Montana State University, Bozeman,
MT 59717, USA}
\author{Yuri Levin}
\affiliation{Physics Department and Columbia Astrophysics Laboratory,
Columbia University, 538  West 120th Street New York, NY 10027, USA}
\affiliation{Center for Computational Astrophysics, Flatiron
  Institute, 162 Fifth Avenue, New York, NY 10010, USA}
\affiliation{Department of Physics and Astronomy, Monash University,
  Clayton
  VIC 3800, AUSTRALIA}

\begin{abstract}

We study pinning and unpinning of superfluid vortices in the inner
crust of a neutron star using 3-dimensional dynamical simulations.
Strong pinning occurs for certain lattice orientations of an
idealized, body-centered cubic lattice, and occurs generally in an
amorphous or impure nuclear lattice.  The pinning force per unit
length is $\sim 10^{16}$ dyn cm$^{-1}$ for a vortex-nucleus
interaction that is repulsive, and $\sim 10^{17}$ dyn cm$^{-1}$ for an
attractive interaction. The pinning force is strong enough to account
for observed spin jumps (glitches). Vortices forced through the
lattice move with a slip-stick character; for a range of superfluid
velocities, the vortex can be in either a cold, pinned state or a hot
unpinned state, with strong excitation of Kelvin waves on the vortex.
This two-state nature of vortex motion sets the stage for large-scale
vortex movement that creates an observable spin glitch.  We argue that
the vortex array is likely to become tangled as a result of repeated
unpinnings and repinnings. We conjecture that during a glitch, the
Kelvin-wave excitation spreads rapidly along the direction of the mean
superfluid vorticity and slower in the direction perpendicular to it,
akin to an anisotropic deflagration.

\end{abstract}

\section{Introduction}

Spin glitches are ubiquitous in neutron stars
(NSs), seen in radio pulsars, x-ray pulsars, magnetars, and
millisecond pulsars.  Typical glitches show a fractional change in
spin rate of $10^{-8}$ to $10^{-6}$.  The spin-up portion of a glitch
has never been fully time resolved; the remarkable 2016 spin glitch of
the Vela pulsar of magnitude $\Delta\nu/\nu=1.4\times 10^{-6}$,
occurred in under 12.6 s \citep{ashton_etal19}, suggestive of a
localized, mechanical process.  The most prolific glitching NS is the
x-ray pulsar J0537-6910, which has suffered 45 glitches in 13 years
\citep{aeka17}. Glitches produce observable changes of the pulsar
magnetosphere \citep{ashton_etal19,palfreyman_2016}; see
\citet{bransgrove2020quake} for discussion.

Glitches are thought to represent sudden coupling between the NS
crust, whose spin rate we directly observe, and the interior
superfluid.  The angular momentum of the superfluid is determined
by the distribution of quantized vortex lines, which are predicted
to pin to the nuclei of the inner crust
\citep{ai75,alpar77,eb88,le91,pvb97,dp2004,dp06,donati_pizzochero2003,abbv07,avogadro2008vortex,link09},
the region of the star with mass density between the nuclear drip
density (average density in baryons, including nuclei) of
$\simeq 2.4\times 10^{-4}$ fm$^{-3}$ ($\simeq 4\times 10^{11}$
\gcc) to about half nuclear density, $\sim 7\times 10^{-2}$
fm$^{-3}$ ($\sim 10^{14}$ \gcc).  As the star spins down under
external torque, vortex pinning fixes the angular velocity of a
portion of the superfluid, storing angular momentum in the region
with pinning, while a hydrodynamic lift force (Magnus force)
gradually increases on the pinned vortices.  As a consequence of
an unidentified instability, a significant fraction of the $\sim 10^{17}$
vortices in the inner crust unpins and moves {\em en masse}.  The
vortices then drag against the crust, driving a spin jump.  A
possible trigger of vortex unpinning is the inevitable structural
relaxation of the crust (a starquake) as the star spins down that
triggers vortex motion, by moving vortices with the matter
\cite{ruderman91c}, or causing vortices to unpin directly
\citep{alpar_etal96}, or locally heating the crust and mobilizing
many vortices through thermal activation \cite{le96}.  Another
possibility is that large-scale unpinning is the result of a
vortex avalanche, wherein an unpinned vortex segment approaches a
pinned vortex segment, causing it to unpin - each unpinned vortex
unpins another vortex and an avalanche develops
\citep{cheng_etal88,wm08,wm11,wmb12,wm13,howitt_etal2020}.

The available angular momentum stored in the superfluid for driving
a spin glitch is determined by the strength to which the vortices
pin to the nuclear lattice.  Pinning occurs as a 
dynamical relaxation process.  A vortex forced through the lattice against
drag will fall into local minima of the interaction potential,
exciting waves on the vortex.  As the waves damp (\eg, through the
excitation of lattice and superfluid phonons), the vortex assumes
a shape that strikes a compromise between the energy gain of falling
into local minima of the lattice potential and the energy cost of
bending the vortex (which increases its length).  Many different
metastable pinning states are possible, so the actual pinned state a
vortex finds will depend on initial conditions.  The strength of
pinning will be determined by: i) the basic interaction between a
vortex and a nucleus, ii) the lattice structure, and the iii)
tension (self-energy) of a vortex.  A vortex will pin to nuclei if
the vortex-nucleus interaction is attractive, and to the interstices
of the lattice if the interaction is repulsive. 

There has been substantial disagreement over the magnitude and sign
of the vortex-nucleus interaction in the inner crust.  Quantum
calculations (using a mean-field Hartree-Fock-Bogoliubov
formulation) show a repulsive interaction with an energy of up to
$\sim 3$ MeV \citep{abbv07,avogadro2008vortex} throughout the inner
crust, while semi-classical calculations (using a local density
approximation) show a repulsive interaction of $\sim 1-2$ MeV below
an average baryon density of $\sim 10^{-2}$ fm$^{-3}$
($\sim 2\times 10^{13}$ \gcc) which turns attractive with a strength
of $\sim 5 $ MeV at higher densities
\citep{pvb97,donati_pizzochero2003,dp2004,dp06}. A step toward
resolving the controversy was made by \citet{wlazlowski_etal16}, using
density functional theory (in principle, an exact approach); they find
that the vortex-nucleus interaction is always repulsive in the average
baryon density range 0.02 fm$^{-3}$ ($3\times 10^{13}$ \gcc) to 0.04
fm$^{-3}$ ($7\times 10^{13}$ \gcc), with a force of $\sim 1$ MeV
fm$^{-1}$ over a range of $\sim 4$ fm, corresponding to an interaction
energy per nucleus of $\sim 4 $ MeV.

The ground state of a single-component Coulomb lattice is a
body-centered-cubic (bcc) structure.  In a bcc lattice, the pinned
vortex will consist of segments that follow the lattice planes,
separated by kinks that traverse the lattice planes, producing some
degree of frustration of the pinned vortex.  As a Magnus force is
applied, the kinks will move.  \citet{jones97} argues that the kinks
move easily, so that pinning to a regular lattice would be very weak.

The inner crust, though, is probably not an ideal bcc
lattice. \citet{jones99,jones01} has shown that thermal fluctuations
will cause the crust to 
solidify with variations in the nuclear charge and lattice defects
(\ie, missing nuclei or mono-vacancies), that destroy long-range order
and most likely make the crust into a glass.\footnote{A highly
  impure crust would have low thermal conductivity, and is disfavored
  by observations of the thermal response of the crust following an
  episode of accretion \citep{bc09}.} \citet{kp14} have shown that,
above the neutron drip density, interstitial neutrons give rise to an
attractive force between nuclei that renders a bcc lattice unstable.
\citet{kp14} suggest that the lattice structure is that of a
superlattice with, \eg, two nuclei per unit cell, analogous to
ferroelectric materials such as BaTiO$_3$, but such a lattice might
never be realized according to the arguments of
\citet{jones99,jones01}.

The first three-dimensional simulations of vortex motion in a
nuclear lattice were performed by \citet{link09} for a nuclear glass
using a classical treatment of the vortex motion and parameterized
potential for the vortex-nucleus interaction. Assuming that the range
of the potential is equal to the lattice spacing, \citet{link09} found
that pinning occurs with equal strength, independent of the sign of
the interaction. All pinning calculations described above show that
the range of the vortex-nucleus interaction is typically smaller than
the lattice spacing, and calculations in this regime are needed, as we
present here. \citet{wlazlowski_etal16} performed three-dimensional
dynamical simulations using density functional theory to obtain the
vortex-nucleus interaction; these calculations considered a vortex
interacting with a single nucleus and so did not address lattice
effects.  Here we use a new numerical algorithm to study vortex
pinning and unpinning with three-dimensional dynamical simulations,
considering different possibilities for the structure of the nuclear
lattice.  We evaluate the vortex pinning force, and identify a
slip-stick character to vortex motion that could play a role in pulsar
glitches.

\section{Vortex Equation of Motion}

We treat a vortex classically as a
string with tension at zero temperature.  Let the vortex shape,
measured with respect to the $z$ axis, be given by the two-dimensional
displacement vector $\ubf(z,t)=\hat{x}u_x+\hat{y}u_y$.  For small
bending angles, so that $|\partial\ubf/\partial z| <<1$, the equation
of motion is \citep{schwarz1977theory,schwarz1978turbulence} \be
T_v\frac{\partial^2\ubf}{\partial
z^2}+\rho_s\kappabf\times\left(\frac{\partial \ubf}{\partial
t}-\vbf_s\right) -\eta\frac{\partial\ubf}{\partial t}+\fbf=0,
\label{eom0} \ee The first term is the force (per unit length) from
bending the vortex; $T_v$ is the vortex tension. The second term is
the Magnus force; $\rho_s$ is the (unentrained) superfluid mass
density, $\kappabf$ is the local vorticity vector, and $\vbf_s$ is the
external superfluid flow velocity. The third term is the drag force
due to motion of the vortex with respect to the nuclear lattice;
$\eta$ is the drag coefficient. The last term $\fbf$ is the force due
to interaction of the vortex with the nuclear lattice.  The effects of
vortex inertia are negligible, so the velocity of any point on the
vortex is determined entirely by the net force at that point and by
the shape of the vortex.  All forces are perpendicular to the vortex.
We solve this equation using a new spectral method described in the
Appendix A.  Given the uncertainties in the vortex-nucleus
interaction, we model the potential parametrically for different
possible lattice structures; details are given in Appendix B. 

The dominant damping process of vortex
motion occurs through the excitation of Kelvin waves on the vortex
\citep{eb92,jones92}, which then exchange energy with lattice
phonons.  While our treatment accounts for the excitation of Kelvin
waves, we do  not treat the excitation of lattice phonons explicitly; the
latter process is treated with the drag term in eq. \ref{eom0}. 
The drag coefficient is typically
$\gamma\equiv\eta/\rho_s\kappa\sim 10^{-3}$ for a vortex that is
moving quickly through the lattice \citep{eb92,jones92}, so Kelvin
waves are underdamped.\footnote{For
low-velocity motion ($v_s<10^3$ \cms), \citet{jones90a} finds
$\gamma\sim 10^{-5}-10^{-4}$. That calculation assumes that
vortices cannot pin, while we find that vortices pin strongly in this
regime.} 
Calculations for $\gamma<<1$ require long run times; 
we fix $\gamma=0.1$ to keep computation times tractable, while
remaining in the expected underdamped regime.  The way in which a
vortex unpins and repins probably has some dependence on $\gamma$,
which we do not study here.\footnote{We find the dependence on
  $\gamma$ to be weak in the interval $0.03<\gamma<0.3$.}

\section{Results of Dynamical Simulations}

\subsection{Bcc lattice}

To illustrate the effects of lattice structure on vortex pinning and
unpinning, we begin with an ideal bcc lattice. For the potential, we
take \be V(\rbf,t)=\frac{E_p}{b}\,
\exp\left[-\frac{b}{\sigma_p}\sum_{i=1}^3\sin^2\left(\pi\hat{e}_i\cdot\frac{\rbf}{b}\right)\right]
\ee where $E_p$ is the interaction energy per nucleus, $b$ is the
lattice spacing, $\sigma_p$ is the length scale of the interaction,
and $\hat{e}_i$ are orthonormal lattice vectors.  The force per unit
length near a nucleus scales as $\sigma_p^{-1}$, distributed over a
segment of the vortex of length $\sigma_p$, so the critical force for
unpinning is nearly independent of $\sigma_p$ for the regime of
interest with $\sigma_p<<b$, scaling as $b^{-2}E_p$.

To determine the threshold for unpinning, we first choose a set of
lattice vectors for the bcc lattice, and then let the vortex relax
dynamically to a pinned configuration with zero external superfluid
flow. We then increase the superfluid flow velocity $v_s$
adiabatically. (If $v_s$ is increased too abruptly, waves are excited
on the vortex that cause it to unpin.)  The vortex remains pinned to a
threshold value of $v_s$. Typically, as $v_s$ is increased, the vortex
undergoes some adjustment that strengthens its position, allowing it
to remain pinned.  As $v_s$ is further increased, the vortex unpins
and starts moving through the lattice, and Kelvin waves are excited on
the vortex. After further increase of $v_s$, the vortex enters a
ballistic regime in which the tension and pinning terms in eq.
\ref{eom0} average to nearly zero.  The solution to eq. \ref{eom0} in
this regime to first order in small $\gamma\equiv\eta/\rho_s\kappa$ is
\be \frac{\partial\ubf}{\partial t}\simeq v_s(\hat{x}-\gamma\hat{y}).
\label{ballistic} \ee The vortex moves at an angle $\simeq\gamma$ with
respect to the direction of the applied flow.  The superfluid velocity
is then slowly decreased to zero; the vortex repins and stops.  A movie
of the motion is shown in Fig. \ref{m1}. 

 \bigskip
\href{https://drive.google.com/file/d/14EBkAhlfLMveMZQ9Rzsql0wFjoK4Mu3J/view?usp=sharing}{Click
  here for animation}

\begin{figure}[H]
\caption{Animation of a vortex forced through an attractive bcc lattice. The
    translational motion has been subtracted. The vortex is 100
    lattice units long along the $z$ axis; the $x$ and $y$ axis have
    been scaled up by a factor of 10. The lattice vectors are
    $\hat{e}_1=[0,0,1]$, 
    $\hat{e}_2=[0,1/2,1/2]$, $\hat{e}_3=[0,-1/2,1/2]$. The force is in
  the $y$ direction, decreasing from zero until
  $t=500$, and then decreasing back to zero by
  $t=1000$. The initial, pinned vortex, obtained by letting the vortex
  damp to the lattice with no external force (not shown), generally
  follows a lattice plane, with jogs and kinks. As the force is
  increased, the vortex eventually becomes unstable and becomes excited
  with helical waves. As the force is reduced, the vortex repins in a
  different configuration than the initial one.}
\label{m1}
\end{figure}

\begin{figure}[H]
\includegraphics[width=\linewidth]{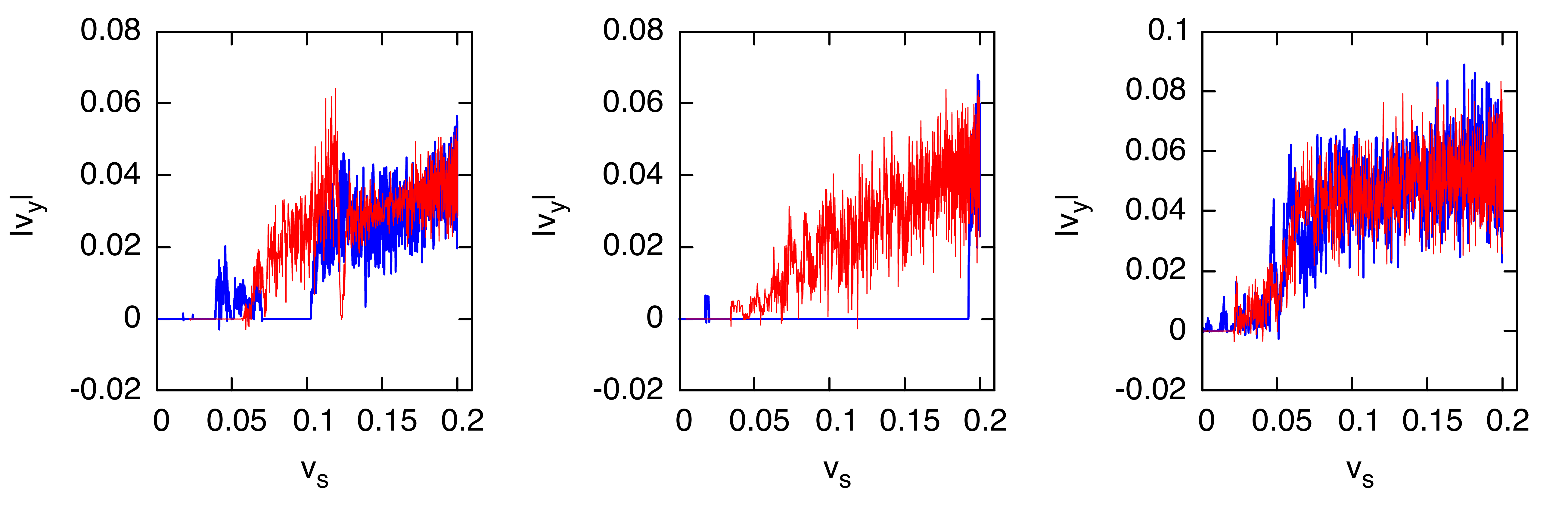}
  \caption{Unpinning/repinning transitions in an attractive bcc lattice
for three arbitrary lattice orientations; shown is vortex speed in the
-$\hat{y}$ direction versus applied superfluid flow velocity $v_s $ in
the $\hat{x}$ direction.  For the blue (red) curves the force is
increasing (decreasing) with time.  In the leftmost figure, the vortex
undergoes some adjustment as the force is increased, before unpinning
suddenly at $v_s\sim 0.1$.  Above this velocity, the vortex enters a
ballistic regime. As the Magnus force is lowered, the vortex does not
repin until $v_s$ is significantly less than the threshold for
unpinning, a strong hysteresis effect.  The rightmost figure shows a
rare example for which there is not a strong unpinning transition.
The parameters are $E_p=-4$ MeV, $b=30$ fm, $\sigma_p=.3b$.  Velocity
units are $T_v/\rho_s\kappa b\sim 2\times 10^8$ cm s$^{-1}$.  }
\label{attractive}
  \end{figure}

We find that pinning depends strongly on the lattice orientation, as
predicted by \citet{jones97}.  For most lattice angles the pinning is
very strong, and for some pinning is virtually nonexistent. The lack
of effective pinning for some lattice angles is due to the presence of
kinks in equilibrium configurations that have very high mobility.
Examples of vortex response are shown in Fig.  \ref{attractive}.
Performing many simulations for different $E_p$, and averaging over
randomly chosen lattice orientations for each value of $E_p$, we find
a pinning force per unit length for an attractive pinning potential of
\be
f_p\simeq\stack{2 \times 10^{-3}}{3\times 10^{17}} \, \left(
\frac{|E_p|}{\mbox{4 MeV}}\right ) \left(\frac{b}{30\mbox{
fm}}\right)^{-2} \quad \stack{\mbox{MeV fm$^{-2}$}}{\mbox{dyn
cm$^{-1}$}}
\label{fp_bcc_attractive}
\ee
By comparison, the strongest possible
pinning occurs for a lattice that is aligned with the vortex. The
pinning force per unit length is then, for the fiducial values above,
$f_p\sim b^{-2}E_p\sim 4\times 10^{-3}\mbox{ MeV fm$^{-2}$}\simeq
7\times 10^{17}\mbox{ dyn cm$^{-1}$}$.  The angle-averaged pinning
force above is nearly this strong, since pinning occurs for most
lattice orientations.  Our estimate for the pinning strength in eq.
\ref{fp_bcc_attractive} is $\sim 300$ times larger than found by
\citet{seveso2016mesoscopic}, and a factor of 3-30 smaller than
estimated by \citet{ai75} with a dimensional argument. 

We note the presence of strong hysteresis in the vortex response to
the force; the vortex typically repins at a much lower value of $v_s$
than the value at which it unpinned.  The reason for this effect is
that the unpinned vortex has a high effective temperature, with
continuous excitation of Kelvin waves that inhibit repinning, while
the immobile (cold) vortex is able to dig into the pinning potential.
The vortex thus has {\em two distinct states for the same applied
force} - one pinned, the other unpinned - and the vortex motion
acquires a {\em slip-stick} character.  This slip-stick character of
vortex motion is generally present, with the repinning value of $v_s$
significantly smaller than the unpinning value, but is absent for
special lattice orientations.  This feature of vortex motion arises
because a vortex, as an extended object, has internal degrees of
freedom.

\begin{figure}[H]
\includegraphics[width=\linewidth]{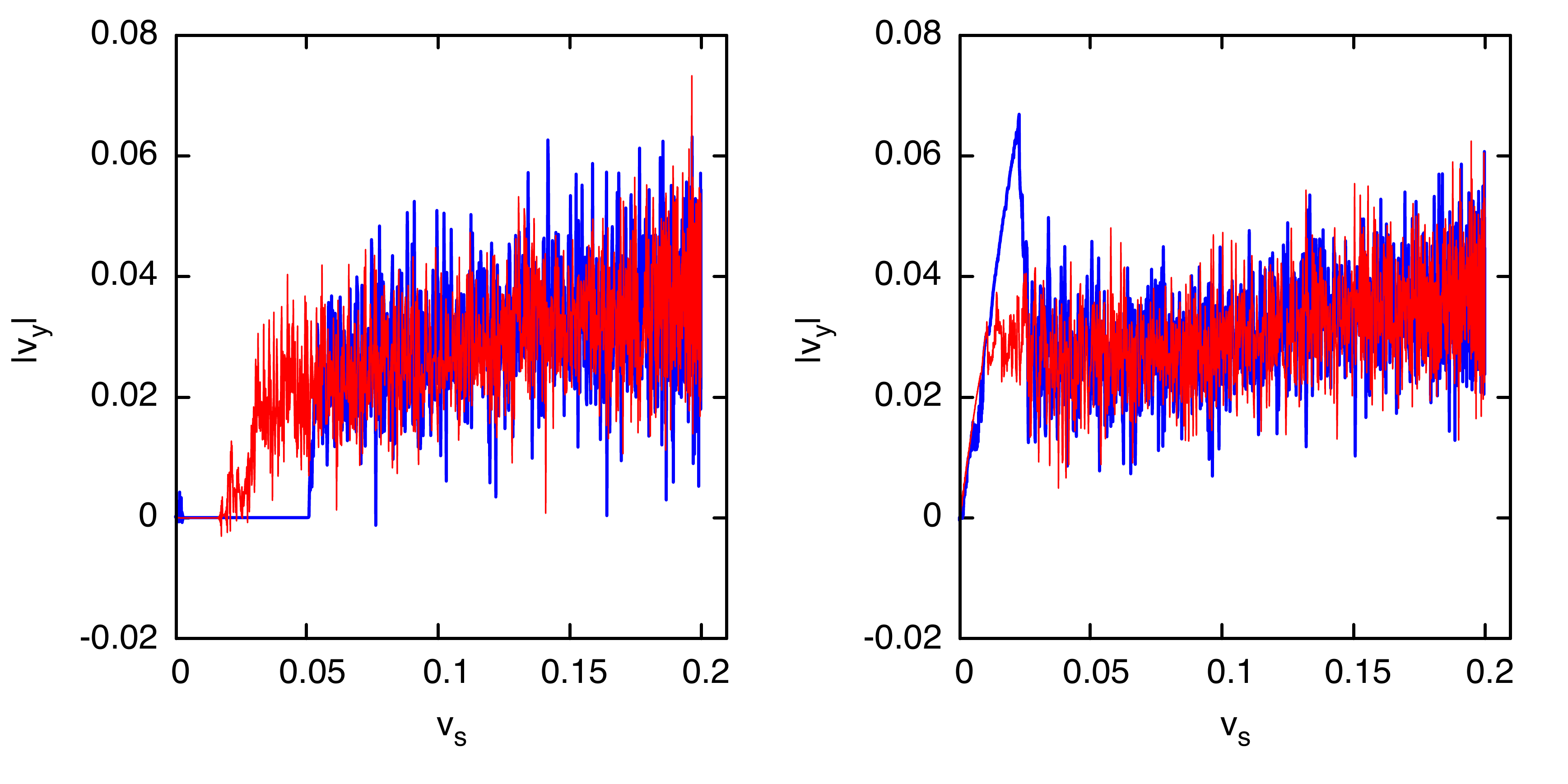}
  \caption{As in Fig.  \ref{attractive}, for a repulsive nuclear
potential.  Typically, there is no pinning transition, as in the right
panel.  For favorable lattice orientations for which there is a sharp
transition, the threshold force is somewhat lower than for an
attractive potential.  }
\label{repulsive}
\end{figure}

For a repulsive nucleus-vortex interaction, as found by
\citet{abbv07} and \citet{avogadro2008vortex}, and
\citep{wlazlowski_etal16}, the vortex pins less strongly to the
interstices of the lattice; see Fig.  \ref{repulsive}.  For most
lattice orientations, there is no pinning; the vortex is able to
wiggle between the nuclei.  Averaging over randomly chosen lattice
orientations, we find: \be f_p\simeq\stack{2 \times 10^{-4}}{3\times
10^{16}} \, \left( \frac{|E_p|}{\mbox{4 MeV}}\right )
\left(\frac{b}{30\mbox{ fm}}\right)^{-2} \quad \stack{\mbox{MeV
fm$^{-2}$}}{\mbox{dyn cm$^{-1}$}}
\label{fp_bcc_repulsive} \ee an order of magnitude smaller than for an
attractive potential, a factor of $\sim 30$ larger than found by
\citet{seveso2016mesoscopic}, and a factor of 30-300 smaller than
estimated by \citet{ai75} with a dimensional argument. 

As long as the lattice is regular, these results are not significantly
changed for different symmetries; the results are similar for a
face-centered cubic (fcc) lattice, or an alloy consisting of an fcc
lattice and a displaced sub-lattice.

\subsection{Impure lattice}

\begin{figure}[H]
\includegraphics[width=\linewidth]{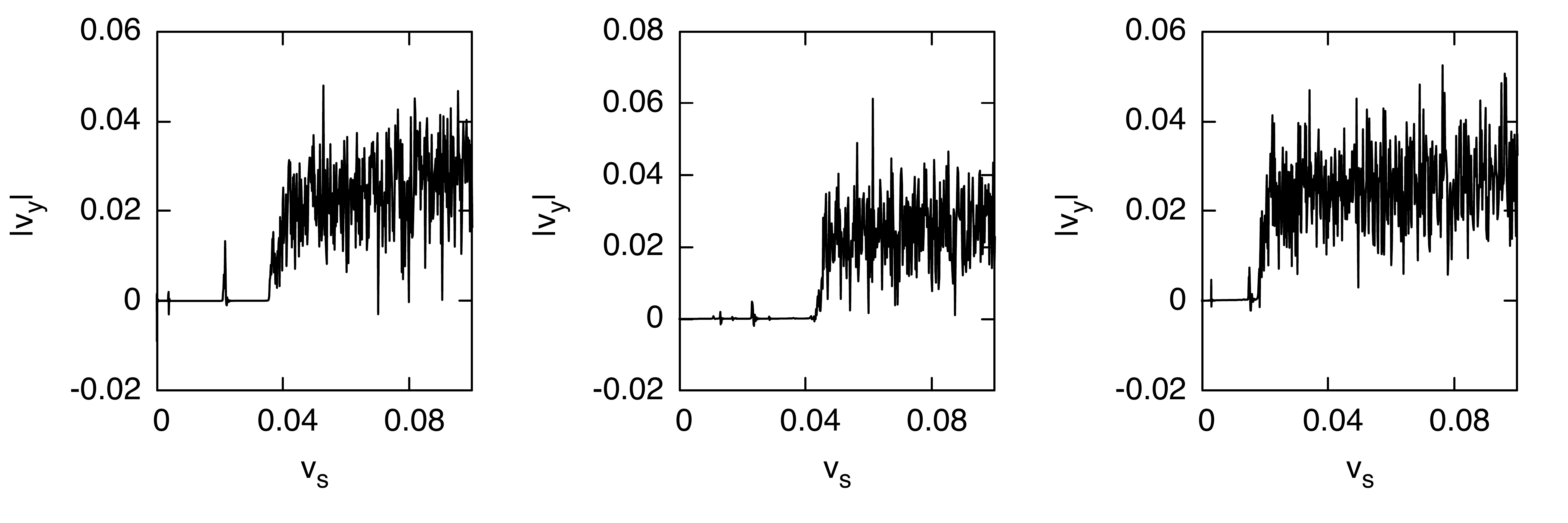}
  \caption{Pinning transitions in a repulsive bcc lattice with a
dilute lattice of randomly placed attractive impurities for three
orientations of the bcc lattice.  Only $v_s$ increasing with time is
shown.  The pinning strength is nearly independent of lattice
orientation, and is determined by the density of impurities.  For
these examples $E_p=2$ MeV, $\sigma_p=.3b$, the impurities are
attractive with strength $-2$ MeV and density 0.02 per unit cell.  }
\label{impure}
\end{figure}

The presence of impurities, such as
dislocations or nuclei with different charge than their neighbors or
mono-vacancies (missing nuclei), can
substantially strengthen pinning for a repulsive lattice.  In a
repulsive lattice, mono-vacancies will attract the vortex.  Fig.
\ref{impure} shows examples for a dilute lattice of randomly
distributed attractive impurities in a repulsive bcc lattice with
$E_p=2$ MeV.  The pinning strength is now largely independent of
lattice orientation as the pinning to impurities is stronger than to
the repulsive lattice.  Impurities have little effect in an attractive
bcc lattice since the pinning to the nuclei is much stronger than
the repulsive case.

\subsection{Nuclear Glass}

\begin{figure}[H]
\centering
  \includegraphics[width=.6\linewidth]{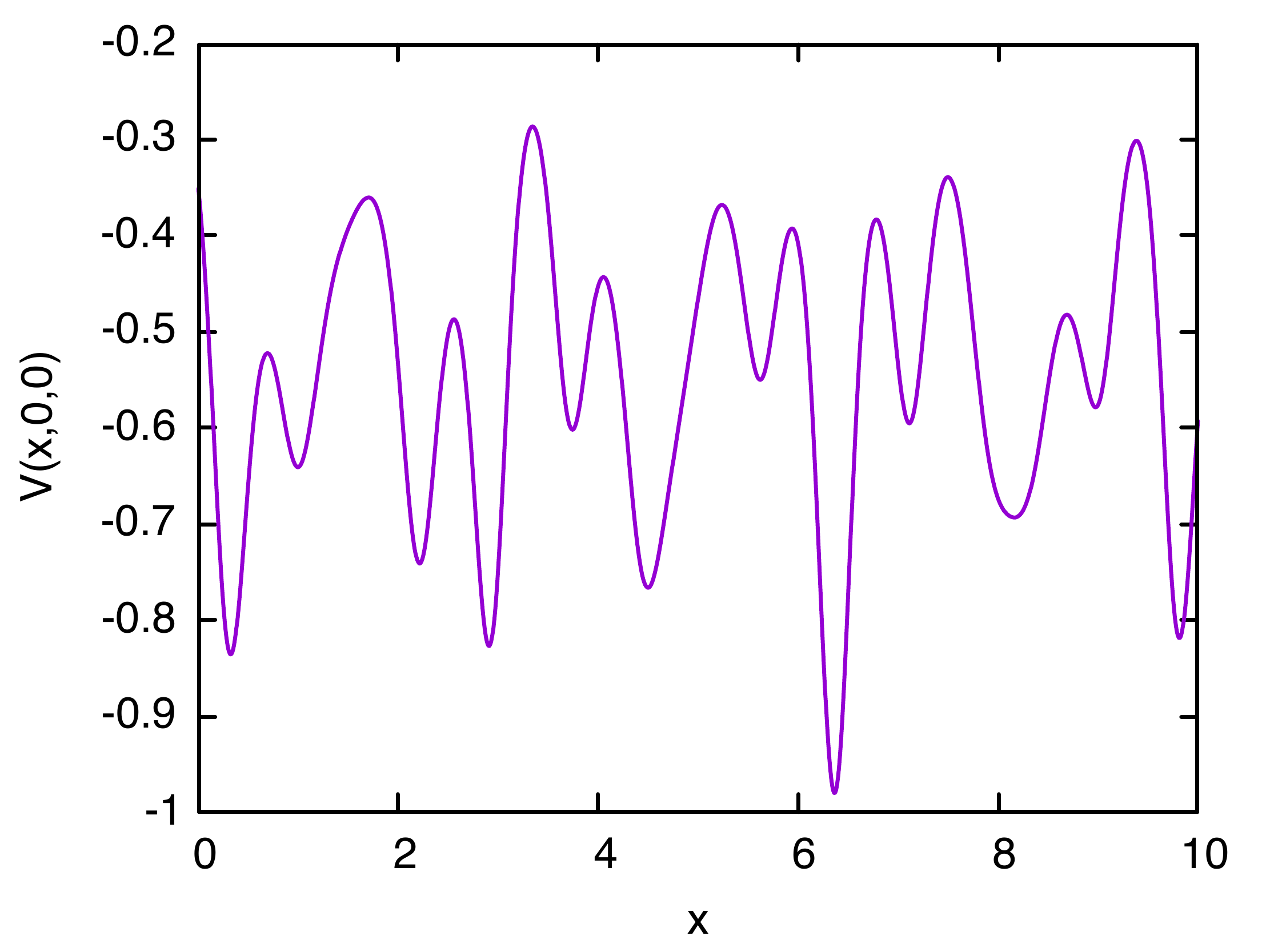}
  \caption{A slice through the random potential.}
\label{random_potential}
\end{figure}

As a model for the potential in a glass, we take 
\[
  V(x,y,z)=\frac{E_p}{b}\,{\rm e}^{-\left(\frac{b}{\sigma_p}\right)^2 [\phi(x)+\phi(y)+\phi(z)]}
\]
with the phases given by random Fourier series of $N$ modes:
\[
  \phi(x) =\frac{1}{N}\sum_{i=1}^N [ c_i \sin ( k_i x + \beta_i)]^2 \qquad
  c_i={\rm rand}(),\ k_i=k_0 [ {\rm rand}()+.5],\ \beta_i=2\pi\, {\rm rand}()
\]
where rand() is the from the interval $[0,1]$, $k_0=\pi/b$, and $b$ is
the mean distance between extrema in the potential.  $\phi(y)$ and $\phi(z)$
are generated in the same way, with their own random numbers.   We
typically use $N=5$.  An example is shown in
Fig.  \ref{random_potential}.

\begin{figure}[H]
\includegraphics[width=\linewidth]{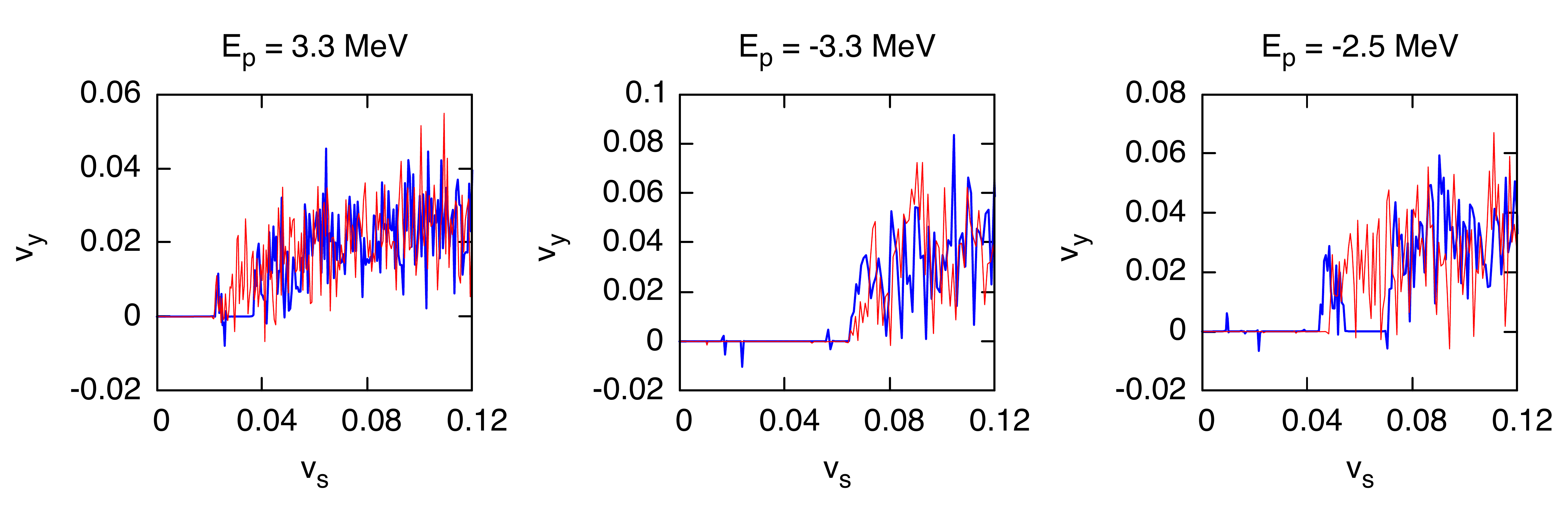}
  \caption{Examples of pinning in a glass.  For these examples, $b=30$
fm.  The hysteresis identified previously is evident in the left and
right plots, but not in the middle one. The first blue spike in the
plot on the left is readjustment of the vortex, not unpinning.  }
\label{glass}
\end{figure}

Pinning to a glass is nearly as strong as
pinning to an attractive lattice for given $E_p$, and is independent
of the sign of $E_p$.  The hysteresis identified for other pinning
situations is usually present, but not always; see Fig. \ref{glass}.
The pinning force per length is nearly equal to that of the attractive
lattice (eq.  \ref{fp_bcc_attractive}).
A movie of vortex motion is shown in Fig. \ref{m2}.

\bigskip
\href{https://drive.google.com/file/d/1J71YUpYAIMeZnXu8Nsc6dDvDM7fCU9AK/view?usp=sharing}{Click
  here for animation}

\begin{figure}[H]
\caption{Animation of a vortex forced through a repulsive glass. The
  coordinates and scalings are as in Fig. \ref{m1}.}
\label{m2}
\end{figure}

\newpage
\section{Discussion}

For an ideal bcc lattice, pinning occurs
for some lattice orientations and is absent for others.  For an
attractive bcc lattice, the angle-averaged pinning force per unit
length is $f_p\sim 10^{-3}$ MeV fm$^{-2}$ $\sim 10^{17}$ dyn
cm$^{-1}$, with a critical flow velocity for unpinning of
$v_c=f_p/(\rho_s\kappa)\sim 10^{-4}c$.  If the vortex-nucleus reaction
is repulsive, as indicated by a number of calculations
\citep{pvb97,donati_pizzochero2003,dp2004,dp06,wlazlowski_etal16},
pinning in a bcc lattice is weakened somewhat to $f_p\sim 10^{-4}$ MeV
fm$^{-2}$ $\sim 10^{16}$ dyn cm$^{-1}$, with a critical flow velocity
for unpinning of $v_c\sim 10^{-5}c$.  If the nuclear lattice consists
of large micro-crystals with different orientations, the pinning
dynamics is likely very complex, with unpinned vortex segments
exciting Kelvin waves that propagate along the vortex and unpin it
elsewhere. Unpinned segments could pin as they move into regions where
the lattice orientation is favorable for pinning.  If the bcc lattice
contains a dilute distribution of attractive impurities (\eg,
mono-vacancies), pinning occurs for any lattice orientation.  If the
nuclear lattice is amorphous, probably a better approximation of the
lattice structure than the bcc idealization \citep{jones01,kp14}, the
pinning is strong, independent of the sign of the vortex-nucleus
interaction, with a pinning force per unit length of $f_p\sim 10^{-3}$
MeV fm$^{-2}$ $\sim 10^{17}$ dyn cm$^{-1}$ (comparable to the
angle-averaged pinning force in an attractive bcc lattice).  Because
a vortex is able to bend as it settles into the pinned state, the
pinning force is much stronger than found by
\citet{seveso2016mesoscopic}, who obtained the force by averaging over
lattice orientations for straight vortices, effectively assuming that
the vortex tension is infinite.

Is the pinning force strong enough to account for large spin glitches?
Suppose the pinned vortices follow the rotation axis of the star of radius
$R$ in the crust of thickness $\Delta R<<R$.  The angular momentum in
the superfluid available to drive a glitch is the excess above that 
for co-rotation of the superfluid with the crust. The
velocity difference between the pinned superfluid and the crust is
$v_s=f_p/\rho_s\kappa$, and the excess angular momentum in the
inertial frame is
\be
\Delta
J_s=\int d^3r\,r\sin\theta\, \rho_s v_s
\simeq 2\pi R^3 \Delta R \,\frac{\bar{f}_p}{\kappa},
\ee
where $\bar{f}_p$ is the mean pinning force in the crust.  At the time of a
glitch, suppose that all of this available angular momentum is given
to the crust plus any other components (\eg, part of the stellar the
core) that are tightly coupled together over the time scale of the
glitch. Let the moment of inertia of the tightly coupled components of
the star be $fI$, where $I$ is the total moment of inertia
of the star and $f$ is the fraction of that moment of inertia that is
tightly coupled.  If the core is not decoupled by a glitch, then
$f\sim 1$, while if the core is completely decoupled by the glitch
$f\sim 10^{-2}$. (In the latter case, observed glitch recovery could
represent response of the core.) The glitch magnitude is given by
$I\Delta\Omega_c=\Delta J_s$, giving
\be
\frac{\Delta\Omega_c}{\Omega_c} \simeq 2 \times 10 ^{-5} \,
f^{-1}\left(\frac{R}{10\mbox{ km}}\right)^4
\left(\frac{\Omega_c}{10^2\mbox{ rad s$^{-1}$}}\right)^{-1}
\left(\frac{I}{10^{45}\mbox{ g cm$^2$}}\right)^{-1} \left(\frac{\Delta
    R/R}{0.05}\right) \left(\frac{\bar{f}_p}{10^{16}\mbox{ dyn
      cm$^{-1}$}}\right)
\ee
Pinning is easily strong enough to account for glitches of magnitude
$\Delta\Omega_c/\Omega_c=10^{-6}$ for any $f\le 1$.

We find the transition from the pinned state to the unpinned state to
be generally very sharp, but it is possible that the transition occurs
over a small range in $v_s$ that we have not resolved in our
simulations at zero temperature.  At finite temperature $T$, the
transition at critical velocity $v_c$ is expected to be smoothed over
a velocity range $\sim (kT/A)\,v_c$, where $A$ is an activation energy
in the range $kT<A<E_p$ \citep{leb93,link14_slippage}\footnote{This
  expectation should be checked with simulations.}.  If the transition
occurs over some range in $v_s$, the slow increase in $v_s$ (over the
stellar spin-down time of $>10^3$ yr), will not produce catastrophic
vortex unpinning; rather, vortex segments will begin to move slowly,
reducing $v_s$ locally, and preventing the vortices from entering the
ballistic regime.  It would appear that a trigger is needed to produce
large-scale vortex motion. One possible trigger is a starquake,
resulting from relaxation of the crust as the star spins down.  A
starquake could cause a portion of the inner crust to move briefly at
a speed comparable to that of shear waves in the matter ($\sim 10^8$ cm s$^{-1}$),
placing vortices on the supercritical side of the curve in Fig. 1. As
the quake ends, $v_s$ is restored to a sub-critical value, but the
vortices are now following the upper, ballistic branch of the
hysteresis curve. The ballistic vortices exert large drag on the crust until
$v_s$ is reduced to nearly zero, and a spin glitch occurs. 

Our numerical experiments demonstrate that unpinning and repinning
are: i) very sensitive to the lattice orientation, and, ii) not
entirely deterministic for a given orientation. It seems inevitable
that repeated unpinnings and repinnings would cause a long vortex line
to deviate significantly from straight.  At each pinning site, the vortex is
bent by a small angle $\theta\propto \left|E_p\right|/T_v$, but the
directions of the kinks are likely to be somewhat random.   The
directions of neighboring vortex segments could differ significantly,
so that the vortex array is tangled to some extent. 

Suppose, during a glitch, a hot, unpinned vortex is driven close to
a cold, pinned vortex.  The repulsive force between the two vortices
could cause the pinned vortex to unpin.  We see in our simulations the
propagation of Kelvin excitations along a single vortex.  We suppose
that, during a glitch, the macroscopic state of the vortex array can
be described by an order parameter - an ``activity'' - that reflects
the strength of the local vortex excitation.  At the start of a
glitch, the activity becomes high in a small region of the crust. The
dispersion in the vortex directions in this region, combined with the
transmission of excitations between the neighboring vortices, will
cause activity to be transported in all directions, and not just along
the mean direction of vorticity. This transport is likely to be
anisotropic, akin to deflagration fronts in many realistic
situations.

Finally, we note that transitions between pinned and unpinned states
depend on the drag force on a moving vortex, an issue deserving
further study.

\begin{acknowledgements}

Y. L. is supported by NSF Grant AST-2009453.  The computations for
this work were performed on the Hyalite High-Performance Computing
Cluster at Montana State University, and on a desktop at the Flatiron
Institute. The Flatiron Institute is a division of the Simons
Foundation, supported through the generosity of Marilyn and Jim
Simons. We thank Andrei Gruzinov for useful discussions.

\end{acknowledgements}

\appendix

\section{Solution Method}

Eq. 1 applies to first order in $\partial\ubf/\partial z$.  In this
limit, variations in the direction of $\kappabf$ do not enter to first
order in the force in directions perpendicular to the vortex, and the
vortex equation of motion becomes \be
T_v\frac{\partial^2\ubf}{\partial z^2} +\rho_s\kappa
\hat{z}\times\left(\frac{\partial \ubf}{\partial t}-\vbf_s\right) -
\eta\frac{\partial\ubf}{\partial t}+\fbf=0.
\label{eom1} \ee Let the unit of length be the lattice spacing $b$,
and define a time unit $\rho_s\kappa b^2/T_v$ ($\sim 10^{-20}$ s in a
neutron star).  Define $\Psi\equiv u_x+iu_y$ and $\tilde{f}\equiv
f_x+if_y$.  Eq.  \ref{eom1} becomes \be \frac{\partial^2\Psi}{\partial
z^2}+(i-\gamma)\frac{\partial\Psi}{\partial
t}-iv_s=-\frac{\tilde{f}}{T_v}
\label{eom} \ee where $\gamma\equiv\eta/\rho_s\kappa$ (dimensionless).
For $\gamma=0$, this equation becomes the one-dimensional
Schr\"odinger equation.

We take the straight vortex to be of length $L$, and the moving vortex
to be defined over the interval $z\in[0,L]$ with free ends.  The
boundary conditions are \be \frac{\partial\Psi}{\partial z}=0\mbox{ at
$z=0$ and $L$}.  \ee

We estimate the vortex tension from hydrodynamics; the vortex tension
is the kinetic energy per unit length in the velocity field produced
by the vortex.  In cylindrical coordinates for a vortex at $r=0$, the
flow velocity is $\vbf_v=\hat{\phi}\,\kappa/2\pi r$.  Treating the
vortex has a hollow core of radius the superfluid coherence length
$\xi$, the tension is (see, \eg, \citet{sonin}).  \be T_v=\half \int
d^2\, \rho_s v_f^2=\frac{1}{4\pi}\rho_s \kappa^2 \ln\frac{L_c}{\xi},
\ee where $L_c$ is an upper cut-off for the integral. As a vortex
moves through the lattice, it becomes bent over a length scale
10-1000$b$; $L_c$ is approximately the bending length in this small
wavelength limit and $\ln L_c/\xi$ is typically $\sim 3$ \footnote{In
the opposite limit of long wavelength, the upper cut-off is comparable
to the inter-vortex spacing, and the logarithmic factor is $\sim 10$
times larger \citep{tk1,tk2,tk2a,tk3,fetter67,bc83}.}. Numerically,
\be T_v\simeq 0.6\left(\frac{\rho_s}{10^{13}\mbox{ \gcc}}\right)\mbox{
MeV fm$^{-1}$}, \ee where a typical value of the unentrained
superfluid mass density was used.  \citet{wlazlowski_etal16} have
argued that the vortex tension could be significantly larger at high
density than the value given by the hydrodynamic estimate.  We take
$T_v=0.6$ MeV fm$^{-1}$ when estimating numerical values. Our
calculations are for fixed $E_p/T_v$, so a larger value of $T_v$
implies a larger value of $E_p$.

We solve eq.  \ref{eom} using a spectral method.  Expand $\Psi(z,t)$
in a Fourier series \be \Psi(z,t)=\sum_{n=0}^\infty a_n(t)\cos k_n z
\qquad k_n=\frac{n\pi}{L}.
\label{decomp} \ee This expansion obeys the free-end boundary
conditions.  The $n=0$ mode corresponds to translation of the vortex.
Substitute eq.  \ref{decomp} into eq.  \ref{eom}, and invoke
orthogonality of $\cos k_nz$ over the domain to obtain an infinite
sequence of coupled, ordinary differential equations \be
(i-\gamma)\dot{a}_n -k_n^2 a_n-iv_s\delta_{n0}=\tilde{f}_n(t) \qquad
n=0..\infty
\label{an} \ee where \be \tilde{f}_n(t)=\frac{2b}{LT_v}\int_0^L dz\,
\tilde{f}(z,t)\cos k_n z \qquad n\ge 1,
\label{fn} \ee and for the translational mode: \be
\tilde{f}_0(t)=\frac{b}{LT_v}\int_0^L dz\, \tilde{f}(z,t) \qquad n=0.
\label{f0} \ee In numerical simulations, we use a finite number of
modes $N_m$ where $N_m$ is determined by the desired spatial
resolution; see below.

To evaluate the force, define a grid in $z$ with spacing $\Delta z$
\be z_i=(i -1) \Delta z \qquad i=1..N_z, \ee so $\Delta z=L/(N_z-1)$.
Eqs.  \ref{fn} and \ref{f0} become \be \tilde{f}_n=\frac{2\Delta
z}{L}\sum_{i=1}^{N_z}M_{ni}F_i
\label{fn_mat} \ee \be \tilde{f}_0=\frac{\Delta
z}{L}\sum_{i=1}^{N_z}M_{0i}F_i
\label{f0_mat} \ee where $M$ is an $N_m\times N_z$ matrix \be
M_{ni}=\cos k_n z_i, \ee and $F_i$ is an $N_z$ dimensional vector \be
F_i=\frac{b\tilde{f}(z_i)}{T_v}.  \ee From eq.  \ref{decomp}, \be
\Psi(z_i,t)=\sum_{n=0}^{N_m} a_n M_{ni}.
\label{Psi} \ee For the force at a point
$\rbf(z,t)=\ubf(z,t)+\hat{z}z$ on the vortex: \be
\fbarbf(\rbf,t)=-\nabla_\perp V(\rbf,t)\equiv -\hat{x}\frac{\partial
V}{\partial x} -\hat{y}\frac{\partial V}{\partial y}.  \ee To solve
the system, we evaluate $\bar{f}_n$ from eqs.  \ref{fn_mat} and
\ref{f0_mat} by matrix multiplication, advance the coupled equations
in eq.  \ref{an} by a time step, and then use eq.  \ref{Psi} to
transform back to coordinate space.

Typically $\Delta z\simeq 10^{-2}b$ gives sufficient resolution, so
$N_z=100L/b$.  The highest wave number should be about $4\pi/b$, so
that $N_m=4L/b$.  For $L=100b$, $N_z=10^4$ and $N_m=400$.

To integrate eq.  \ref{an}, we use best to use fifth-order Runge-Kutta
with adaptive step size to control error.


\end{document}